\documentclass[aps,twocolumn,english,superscriptaddress,citeautoscript,preprintnumbers,amsmath,amssymb,floatfix,footinbib]{revtex4}
\usepackage{hyperref}
\hypersetup{colorlinks=true,linkcolor=red,citecolor=blue}
\usepackage{amsmath}
\usepackage{graphicx}
\usepackage{dcolumn}
\usepackage{float}

\begin{document}
\title{Anisotropic magnetotransport and magnetic phase diagrams of the antiferromagnetic heavy-fermion superconductor Ce$_3$PdIn$_{11}$ }
\author{Debarchan Das}
\affiliation{Institute of Low Temperature and Structure Research, Polish Academy of Sciences, Wroc{\l}aw 50-950, Poland}
\author{Daniel Gnida}
\affiliation{Institute of Low Temperature and Structure Research, Polish Academy of Sciences, Wroc{\l}aw 50-950, Poland}
\author{Dariusz Kaczorowski}
\email{d.kaczorowski@int.pan.wroc.pl}
\affiliation{Institute of Low Temperature and Structure Research, Polish Academy of Sciences, Wroc{\l}aw 50-950, Poland}

\begin{abstract}
We report the results of our detailed magnetotransport studies on single crystals of the antiferromagnetic heavy-fermion superconductor Ce$_3$PdIn$_{11}$. Electrical resistivity measurements, carried out in different magnetic field orientations with respect to the crystallographic axes, were mainly aimed to amass further insight about the magnetically ordered state. The results manifest a clear metamagnetic transitions in the ordered phase when the applied field is parallel to the tetragonal $c$ axis, while no similar features are seen for the transverse direction, i.e., with the field confined within the $ab$ plane. This finding elucidates the fact that the $c$ axis is the easy magnetic direction in this system. Based on the electrical transport and heat capacity data obtained for Ce$_3$PdIn$_{11}$, magnetic field -- temperature phase diagrams were constructed, which elucidate fairly enigmatic behaviors in this material featuring the existence of both second- and first-order magnetic phase transitions.
	
\end{abstract}

\date{\today}

\maketitle

\section{Introduction}

Since the seminal discovery of heavy-fermion (HF) superconductivity in CeCu$_2$Si$_2$ \cite{Steglich1}, Ce-based intermetallic compounds have been a subject of intensive research activity. As a result, diverse ground state properties have been discovered featuring HF behavior, non-Fermi liquid features, magnetic ordering, superconductivity, quantum criticality, etc. \cite{Stewart1, Stewart2, Steglich2, Steglich3, Gegenwart}. Particular attention is focused on quantum critical HF superconductors where  significant enhancement in quasiparticle effective masses is associated with formation of superconducting condensate in the vicinity of magnetic instability \cite{Si,Kuchler, Agterberg}. Unlike conventional superconductors, where Cooper pairing is mediated by phonons, in quantum critical ones it is arguably believed to be driven by magnetic fluctuations \cite{Mathur,Stockert}. This intriguing concept steers the condensed matter research community in the quest to look for novel HF superconductors and thoroughly explore their puzzling physics. In this framework, the homologous series of Ce$_n$T$_m$In$_{3n+2m}$ compounds (T stands for a $d$-electron transition metal) turns out to be an extremely important system as it encompasses a large variety of fascinating materials including CeCoIn$_5$ \cite{Petrovic}, CeRhIn$_5$ \cite{Park}, CeIrIn$_5$ \cite{Shang}, Ce$_2$PdIn$_8$ \cite{Kaczorowski, Kaczorowski2, Kaczorowski3}, Ce$_2$CoIn$_8$ \cite{Chen}. Two other remarkable representatives of the same family of ternaries are the indides Ce$_3$TIn$_{11}$ (T = Pt and Pd), which exhibit the coexistence of HF superconductivity and long-range antiferromagnetic (AFM) ordering \cite{Prokle, Custers2, Kratochv,Das}, and thus hosting the most intriguing physics witnessed in iron pnictides \cite{Kamihara, Johnston,Jeevan, Paramanik, Zapf}. Interestingly, detailed investigations on Ce$_3$PtIn$_{11}$  led to the discovery of several interesting phenomena, like quantum criticality near hydrostatic pressure of about $p_{\rm c}$ = 1.3~GPa \cite{Custers2}, or complex magnetic field -- temperature phase diagram \cite{Das}. Thus, Ce$_3$PtIn$_{11}$ and Ce$_3$PdIn$_{11}$ set a new playground for comprehensive studies on the interplay between magnetism and superconductivity in Ce-based HF systems. The complex nature of hybridization between cerium $4f$ and transition metal $3d$ orbitals is responsible for governing such captivating ground state properties. In this context, it is worth to refer to another Ce-based material CePt$_3$Si, which also exhibits superconductivity in the AFM ordered state \cite{Bauer1,Kaur, Bauer2}.

Similar to its Pt-bearing counterpart \cite{Prokle, Custers2,Das}, Ce$_3$PdIn$_{11}$ crystallizes with a tetragonal unit cell (space group P$4/~mmm$), which features two inequivalent Ce sites \cite{Tursina}, and exhibits two successive AFM phase transitions at $T_{\rm {N1}}$ = 1.67~K and $T_{\rm {N2}}$ = 1.53~K, followed by a superconducting transition at $T_{\rm c}$ = 0.42~K \cite{Kratochv}. A comprehensive study by M. Kratochv\'{i}lov$\acute{\textup{a}}$ $et~ al.$ \cite{Kratochv}  revealed a fascinating magnetic phase diagram where two AFM orders were found to merge in a critical magnetic field of about 3~T, applied along the tetragonal $c$ axis, followed by re-splitting into two AFM transitions in external fields stronger than 4~T. Most interestingly, pronounced sharp anomalies observed in the heat capacity data taken above the critical field were associated with first--order type phase transition, which demanded further experimental confirmation. Our recent work on the heat capacity of Ce$_3$PtIn$_{11}$ revealed similar phase diagram \cite{Das}. In addition, by studying magnetotransport in the latter compound, we found the existence of clear metamagnetic transition (MMT). This finding motivated us to direct our on-going systematic research on the Ce$_n$T$_m$In$_{3n+2m}$ indides towards exploration of the actual character of the AFM state in Ce$_3$PdIn$_{11}$. The goal of the present study was mainly twofold: (i) verification of the first--order nature of the AFM transitions in strong magnetic fields, as suggested by M. Kratochv\'{i}lov$\acute{\textup{a}}$ $et~ al. $ \cite{Kratochv} and (ii) checking the possible presence of MMT in Ce$_3$PdIn$_{11}$. For this purpose, we performed heat capacity and electrical resistivity measurements on single-crystalline specimens in external magnetic fields applied along the two principal crystallographic orientations, namely parallel and perpendicular to the tetragonal $c$ axis. The heat capacity data have not revealed any noticeable hysteresis between cooling and heating cycles, thus severely challenging the possibility of the first--order character of the AFM transitions investigated. In turn, the magnetotransport results obtained in the configuration $\mu_0H \parallel c$-axis have clearly shown metamagnetic transition at temperatures below $T_{\rm {N2}}$, while no such signature of MMT has been found for $\mu_0H\perp$ $c$-axis. From the experimental data obtained for Ce$_3$PdIn$_{11}$, magnetic phase diagrams have been constructed, which esentially appear fairly similar to those derived in Ref.\cite{Kratochv}.

\section{Experimental details}

Single crystals of Ce$_3$PdIn$_{11}$ were grown from In flux following the method outlined by M. Kratochv\'{i}lov$\acute{\textup{a}}$ $et~ al.$ \cite{Kratochv}. Phase purity and homogeneity of the crystals were examined by energy-dispersive X-ray (EDX) analysis using a FEI scanning electron microscope equipped with an EDAX PV9800 microanalyzer. Their crystal structure was determined by x-ray diffraction (XRD) on a KUMA Diffraction KM-4 four-circle diffractometer equipped with a CCD camera, using graphite-monochromatized Mo-K$\alpha$ radiation. The EDX and XRD data indicated good quality of the obtained materials with the crystal structure reported before by Tursina $et~al.$ \cite{Tursina}.

The electrical resistivity measurements were carried out over the temperature interval 0.4 -- 300~K and in magnetic fields up to 9~T using a standard ac four-probe technique implemented in a Quantum Design PPMS platform. The heat capacity was measured from 0.35 to 20~K in magnetic fields up to 9~T by relaxation method employing the same PPMS equipment.

\section{Results and Discussions}

\subsection{Zero-field heat capacity and electrical resistivity measurements}

In order to characterize the physical properties of the obtained single crystals of Ce$_3$PdIn$_{11}$, temperature variations of their specific heat ($C$) and electrical resistivity ($\rho$) were determined. 

Fig. 1a presents the low-temperature $C(T)$ data, which manifests two successive AFM transitions at $T_{\rm {N1}}$ = 1.68~K and $T_{\rm {N2}}$ = 1.56~K, in good agreement with the previous study \cite{Kratochv}. At $T_{\rm c}$ = 0.58~K, there occurs another pronounced anomaly in $C(T)$ that can be associated with the onset of the superconducting state, though the critical temperature is slightly higher than that reported in the literature ($T_{\rm c}$ = 0.42~K \cite{Kratochv}). It seems possible that the sample studied by us contained tiny amount of the superconducting Ce$_2$PdIn$_8$ phase ($T_{\rm c}$ = 0.68~K \cite{Kaczorowski}) sandwiched between crystalline slabs of Ce$_3$PdIn$_{11}$. In this same context, it should be stressed that the $C(T)$ dependence showed no singularity at 10~K (not shown), hence proving that the sample investigated was free of any CeIn$_3$ contamination (see the discussion in Refs. \cite{Uhlirova, Kaczorowski4, Kaczorowski5}).

 \begin{figure}[htb!]
	\includegraphics[width=8.5cm, keepaspectratio]{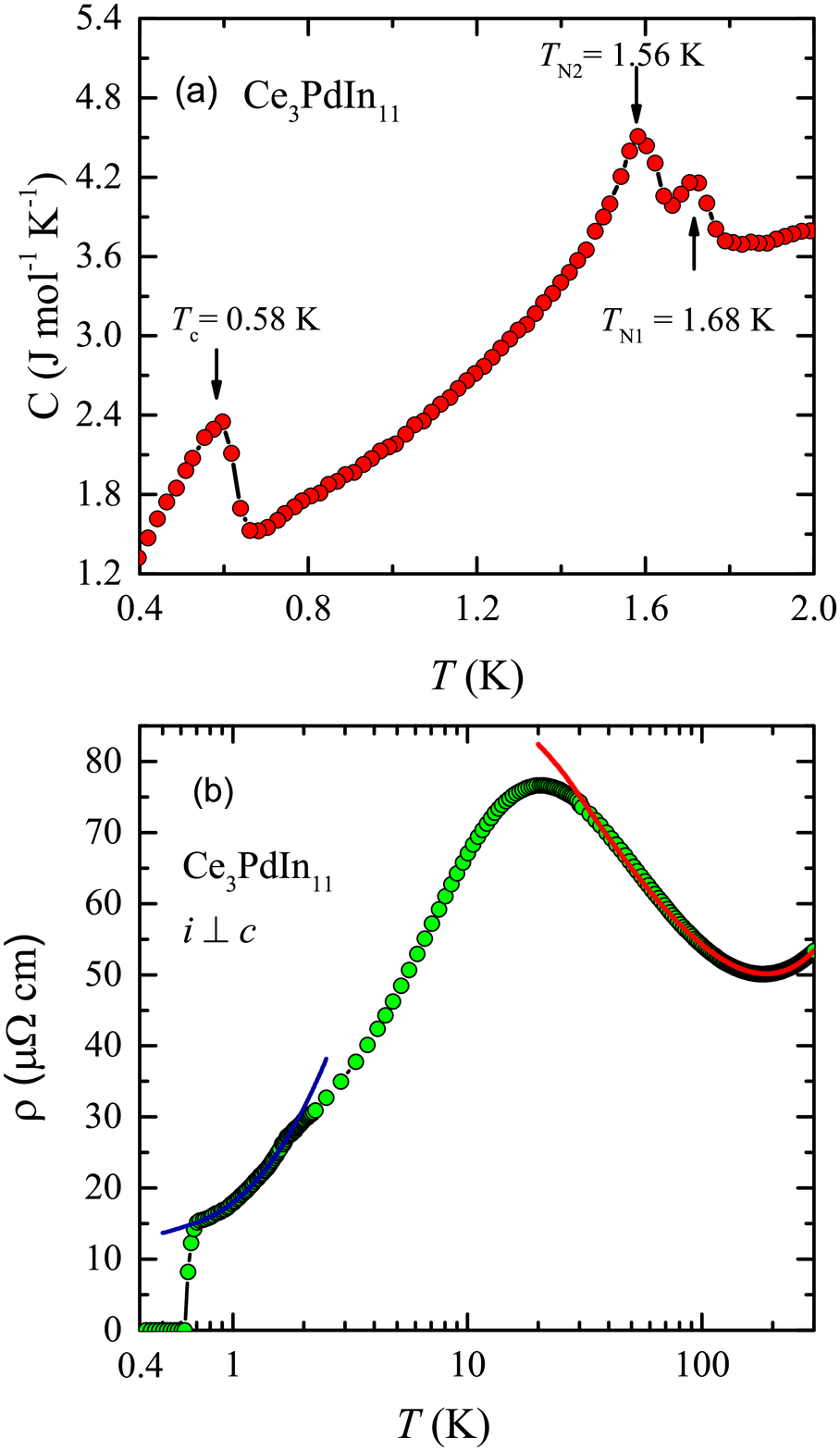}
	\caption{\label{fig:Heat Capacity RT}(Color online)(a) Low-temperature dependence of the specific heat of single crystalline Ce$_3$PdIn$_{11}$. (b) Temperature variation of the electrical resistivity of single crystalline Ce$_3$PdIn$_{11}$ measured with electric current flowing within the tetragonal $ab$ plane, i.e. perpendicular to the $c$ axis.}
\end{figure}

Fig. 1b displays the electrical resistivity of Ce$_3$PdIn$_{11}$ measured as a function of temperature with electric current flowing within the basal $ab$ plane of the tetragonal unit cell. Above about 30~K, $\rho (T)$ can be described by the formula (note the red solid line in Fig. 1b)

\begin{equation}
\rho(T) = \rho_0 + \rho_0^\infty + c_{\rm {ph}}T + c_{\rm {K}} \ln T       ,
\end{equation}

\noindent where $\rho_0$ represents the residual resistivity due to scattering conduction electrons on crystal imperfections, $\rho_0^{\infty}$ is the spin-disorder component due to elastic scattering on cerium magnetic moments in the paramagnetic state (here, for simplicity, crystalline electric field effect is neglected), the third term accounts for electron-phonon scattering expressed as high-temperature approximation of the Bloch-Gr\"{u}neissen function, while the forth term describes Kondo type spin-flip scattering processes. Fitting Eq.~1 to the experimental data yielded the parameters: $\rho_0 + \rho_0^{\infty} = $156.8(5)~$\mu \Omega$cm, $c_{\rm {ph}}$ = 0.134(1) $\mu \Omega$cm/K and $c_{\rm {K}} $= -25.2(1) $\mu \Omega$cm. The notably large value of $c_{\rm {K}} $ manifests strong Kondo interactions in the compound investigated.

Near 20~K, $\rho(T)$ shows a broad maximum (cf. Fig. 1b) that can be associated with a crossover from incoherent to coherent Kondo regimes, typical for Ce-based Kondo lattices. At lower temperatures, a distinct kink in $\rho (T)$ is seen, which develops due to rapid reduction in the spin-disorder scattering in the AFM state. The critical temperature, defined by the maximum in the temperature dependence of the derivative d$\rho$/d$T$, amounts to 1.56~K, and thus it is equal to $T_{\rm {N2}}$ determined from the heat capacity. It is worth noting that the onset of AFM was hardly detectable in the $\rho(T)$ data of Ce$_3$PdIn$_{11}$ reported before \cite{Kratochv,Tursina}. Its clear observation in the present research corroborates the good quality of the samples investigated. As can be inferred from Fig. 1b, below $T_c$ $\sim$~0.58~K, the electrical resistivity of the measured crystal drops to zero, as expected for the superconducting state.

In order to quantitatively analyze the $\rho (T)$ data in the AFM state, a gapped spin-wave approach was applied. As marked in Fig. 1b by the blue solid line, in the temperature interval
$T_c~<~T~<~T_{\rm {N2}}$, the experimental results can be well approximated by the formula \cite{Fontes}

\begin{equation}
\begin{gathered}
\rho(T) = \rho_0 + A~T^2 + b~\Delta_{\rm {SW}}^2~\sqrt{\frac{T}{\Delta_{\rm {SW}}}}~\textup{exp}\left(-\frac{\Delta_{\rm {SW}}}{T}\right)\\
\times~\left[1+\frac{2}{3}\frac{\Delta_{\rm {SW}}}{T}+\frac{2}{15}\left(\frac{\Delta_{\rm {SW}}}{T}\right)^2\right] ,
\end{gathered}
\label{eq:2}
\end{equation}

\noindent where the $T^2$ term describes the Fermi liquid contribution, and the third term accounts for  scattering conduction electrons on AFM magnons with an energy gap $\Delta_{SW}$ in their excitations  spectrum (here, at such low temperatures, it is assumed that phonons are almost frozen, and hence their   contribution to $\rho(T)$ can be neglected). The coefficient $b$ in this expression is related to the spin-wave stiffness D as $b~ \sim D^{-\frac{3}{2}}$ \cite{Fontes}. The least squared fitting of Eq. 2 to the experimental data yielded the parameters: $\rho_0$ = 13.6(4)~$\mu\Omega$cm, $A$ =1.4(8) $\mu\Omega$~cm~K$^{-2}$, $b$ = 2.9(5) $\mu\Omega$~cm~K$^{-2}$ and $\Delta_{\rm {SW}}$ = 3.2(2)~K. It is worth noting that the so-obtained value of $\Delta_{\rm {SW}}$ is close to that estimated by M. Kratochv\'{i}lov$\acute{\textup{a}}$ $et~ al.$ \cite{Kratochv} from the heat capacity data ($\Delta_{\rm {SW}}$ = 2.74~K). Combining the results obtained from fitting $\rho (T)$ with Eq.~1 and Eq.~2, one can estimate the spin-disorder resistivity in Ce$_3$PdIn$_{11}$ to be $\rho_0^\infty$ = 143.2~$\mu\Omega$cm.

\subsection{Heat capacity and electrical transport measurements in different magnetic field orientations}
\begin{figure}[htb!]
	\includegraphics[width=8.5cm, keepaspectratio]{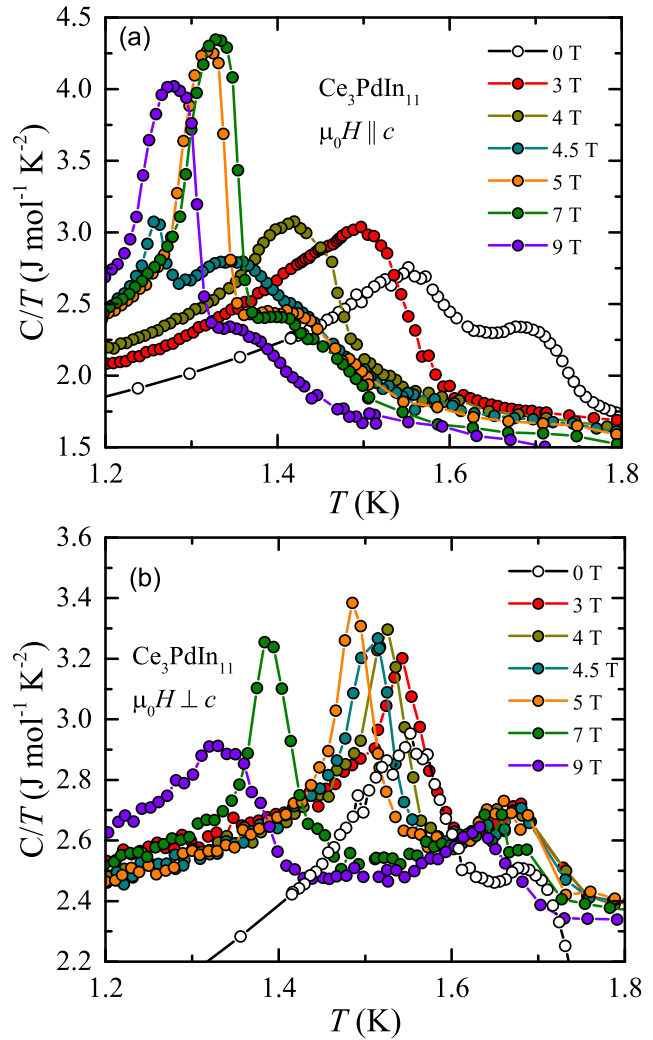}
	\caption{\label{fig:C(T)_in_fields}(Color online) Low-temperature heat capacity of single crystalline Ce$_3$PdIn$_{11}$ measured in different magnetic fields applied (a) along and (b) perpendicular to the crystallographic $c$ axis. For better clarity, the ratio specific heat over temperature is plotted.}
\end{figure}

Fig.~2a depicts the temperature dependence of $C(T)/T$ measured on a single crystal of Ce$_3$PdIn$_{11}$ in magnetic fields applied along the crystallographic $c$ axis. It is quite evident from the figure that the two transitions at $T_{\rm {N1}}$ and  $T_{\rm {N2}}$ initially shift towards lower temperatures with increasing magnetic field, which corroborates the AFM nature of the ordering. In a field stronger than 2.5~T, these two lambda-type anomalies merge into a single sharp feature (labeled hereafter as $T_{\rm {M}}$). The latter singularity moves to lower temperature with the field strength raising up to 4~T. In $\mu_0H\geq$4.5~T, this anomaly again splits into two separate features (labeled as $T_{\rm {M1}}$ and  $T_{\rm {M2}}$ for higher and lower temperature anomalies respectively). Notably, for 4.5~T $\leq\mu_0H\leq$ 9~T, as the field is increased, the height of the peak $T_{\rm {M1}}$ systematically decreases whereas the height of $T_{\rm {M2}}$ dramatically sharpens along with increase in the absolute value of the peak. The overall behavior of these anomalies is similar to that reported by  M. Kratochv\'{i}lov$\acute{\textup{a}}$ $et~ al.$ \cite{Kratochv}. The later authors also investigated the field dependence of $T_{\rm {N1}}$ and  $T_{\rm {N2}}$ in the configuration $\mu_0H \perp$ $c$-axis, yet their heat capacity measurements were limited to 5~T only. Up to that field strength, the transitions at $T_{\rm {N1}}$ and $T_{\rm {N2}}$ remained separated with the value of $T_{\rm {N1}}$ decreasing with increasing $\mu_0H$, while that of $T_{\rm {N2}}$ being almost constant. As shown in Fig. 2b, similar behavior persists up to $\mu_0H$ = 9~T. It is worth pointing out that the heat capacity peak associated with $T_{\rm {N2}}$ sharpens with increasing field up to 7~T (see Fig 2b).  

\begin{figure}[htb!]
	\includegraphics[width=8.5cm, keepaspectratio]{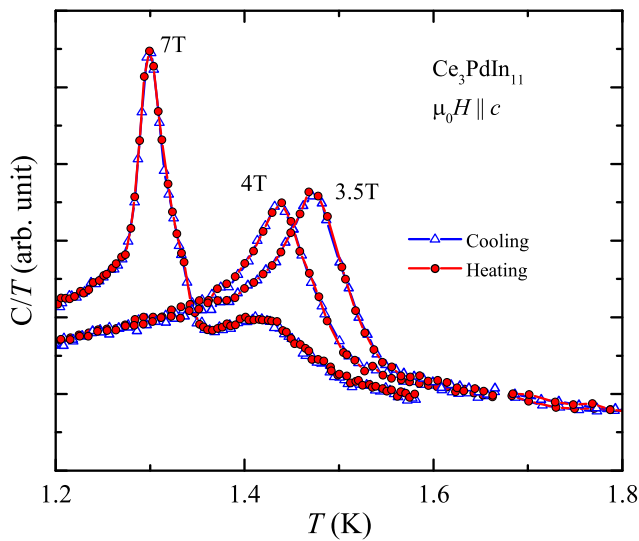}
	\caption{\label{fig:Hyst_C}(Color online) Heat capacity of single-crystalline Ce$_3$PdIn$_{11}$ measured in various magnetic fields applied parallel to the $c$ axis on cooling and heating the specimen.}
\end{figure}

Tempted by the fact that observed sharp anomalies in $C(T)$ can be associated with the first-order type transition  \cite{Kratochv}, the heat capacity measurements were carried out (on a different single crystalline piece) in cooling and heating regimes. As it is clear from Fig. 3, no noticeable hysteresis was observed, as might be expected for latent heat effect. This finding challenges the scenario of first--order nature of the AFM transitions occurring in Ce$_3$PdIn$_{11}$ in strong magnetic fields.

\begin{figure}[htb!]
	\includegraphics[width=8.5cm, keepaspectratio]{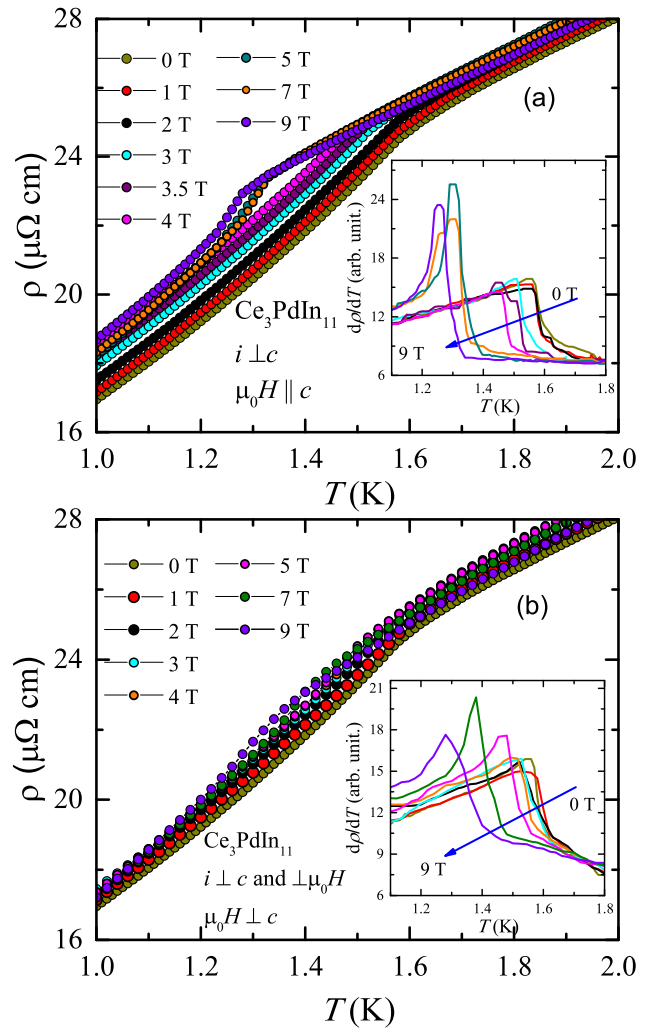}
	\caption{\label{fig:resistivity_in_field}(Color online) Low-temperature electrical resistivity of single crystalline Ce$_3$PdIn$_{11}$ measured with electrical current flowing within the tetragonal plane in different magnetic fields applied (always perpendicular to the current) (a) along and (b) perpendicular to the crystallographic $c$ axis. Insets: temperature dependencies of the temperature derivative of the resistivity measured as in the main panels.}
\end{figure}

In order to gain further insight on the AFM ordering in Ce$_3$PdIn$_{11}$, the electrical resistivity was measured on the single-crystalline specimens in various external magnetic fields applied in two different directions with respect to the crystallographic $c$ axis. As can be inferred from Fig. 4, regardless the field orientation and the field magnitude, one observes a single feature in $\rho(T)$, leading to a maximum in $d\rho/dT$ vs $T$, which defines the AFM transition (see the insets to Fig. 4). With increasing field, the anomaly in $\rho (T)$ at the ordering temperature systematically shifts to lower temperatures. In the case of $\mu_0H\parallel$ $c$-axis, it becomes very sharp in fields $\mu_0H>$ 4~T (note a rapid drop in the resistivity and a distinct peak in $d\rho/dT (T)$), in concert with the character of the peaks in $C(T)$. For $\mu_0H\perp$ $c$-axis, the kink in $\rho (T)$ is less pronounced even in the strong fields (cf. Fig. 4b and its inset), again in line with the heat capacity data. Clearly, the effect of external magnetic field on the magnetic behavior in Ce$_3$PdIn$_{11}$ is highly anisotropic, and clarification of its actual microscopic origin is a tempting issue that calls for further dedicated investigations.

\subsection{Transverse magnetoresistance}

For an AFM system, magnetoresistance (MR) measurements can provide valuable information about conduction electrons scattering processes and its high-field magnetic states. Defining MR = $\frac{\rho(\mu_0H)-\rho(0)}{\rho(0)}$, the MR data were collected for Ce$_3$PdIn$_{11}$ with $\mu_0H\parallel$ $c$-axis and $\mu_0H \perp$ $c$-axis, and the electric current flowing within the crystallographic $ab$ plane (field and current directions were always perpendicular to each other). As shown in Fig. 5a and c, regardless the experimental geometry, the transverse MR of Ce$_3$PdIn$_{11}$ is positive in both the ordered and paramagnetic regions. 

\begin{figure}[htb!]
	\includegraphics[width=8.5cm, keepaspectratio]{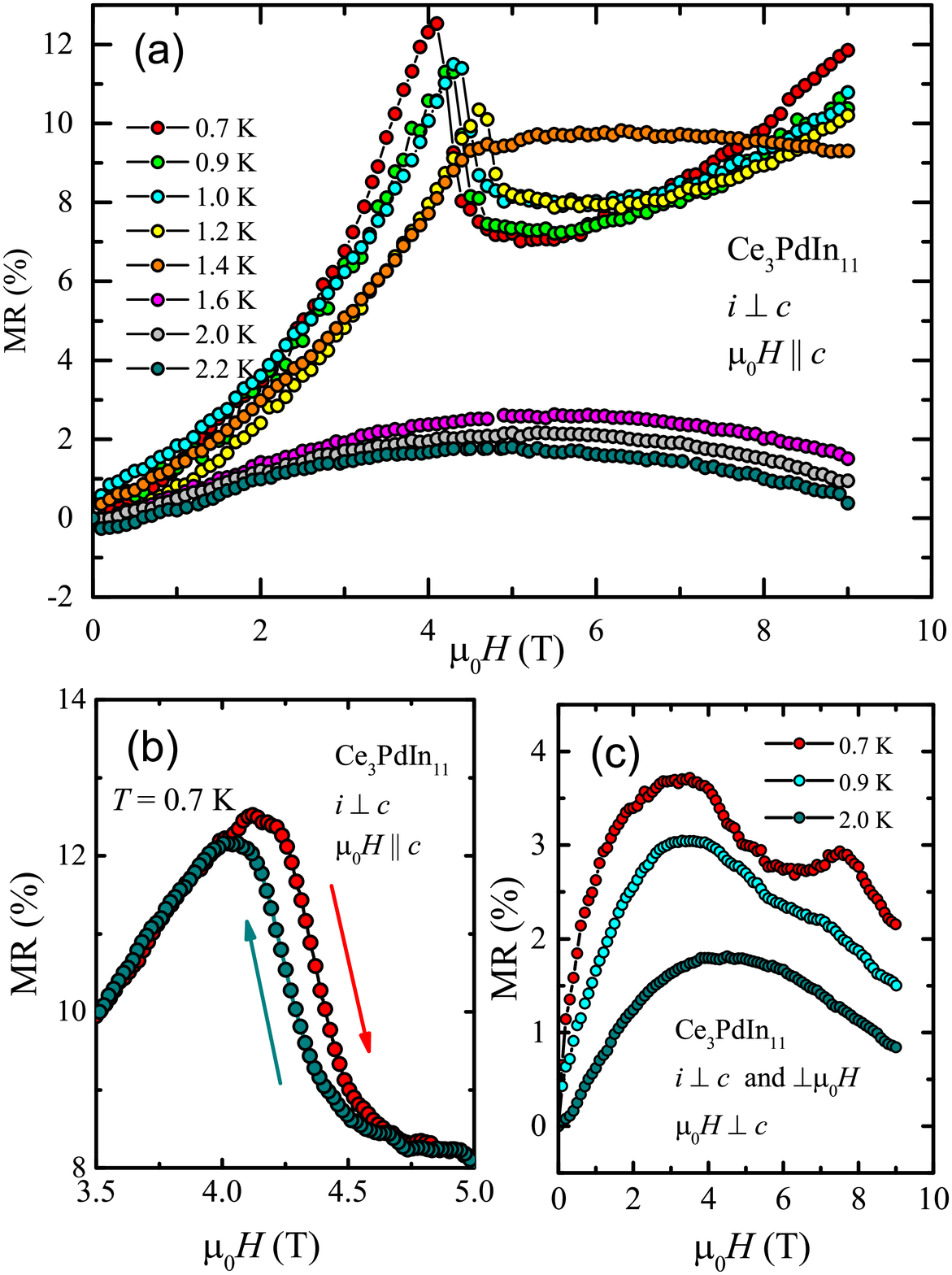}
	\caption{\label{fig:Magnetoresistance}(Color online) (a) Magnetic field dependencies of the transverse magnetoresistance of single-crystalline Ce$_3$PdIn$_{11}$ measured at several temperatures with electrical current flowing within the tetragonal $ab$ plane and magnetic field applied along the crystallographic $c$ axis. (b) Magnetic field variations of the transverse magnetoresistance of Ce$_3$PdIn$_{11}$scan measured at 0.7~K as in panel (a) with increasing and decreasing field strength. (c) Magnetic field dependencies of the transverse magnetoresistance of single-crystalline Ce$_3$PdIn$_{11}$ measured at several temperatures with electrical current flowing within the tetragonal $ab$ plane and magnetic field applied perpendicular to the crystallographic $c$ axis.}
\end{figure}

In the case of $\mu_0H\parallel$ $c$-axis, far below $T_{\rm {N1}}$, MR increases with ramping field up to a critical field $\mu_0H_{\rm c}$, at which a pronounced peak is seen (see Fig. 5a). While the observed positive value of MR is consistent with the AFM type of ordering, the MR singularity at $\mu_0H_{\rm c}$ can be interpreted as an indication of metamagnetic-like phase transition. Remarkably, MMT shows a non-monotonic temperature dependence as it initially moves towards stronger fields on rising $T$ up to 1.2~K, but with further temperature increase, the value of $\mu_0H_{\rm c}$ slightly decreases. Close to $T_{\rm {N2}}$, this feature in MR is quite broadened and then disappears at higher temperatures. Surprisingly, at 1.6~K the overall behavior of MR changes abruptly which certainly indicates a different type of scattering mechanism in the system. In order to examine the naturally expected first--order character of MMT, the MR data were collected at $T$ = 0.7~K on sweeping the magnetic field strength to larger and lower values. As evident from from Fig. 5b, MR isotherms show a clear hysteresis around MMT, which confirms the first--order type of this transition. 

Fig. 5c presents the transverse MR data collected in the configuration $\mu_0H\perp$ $c$-axis. The overall behavior of MR is distinctly different from that observed in other field orientation. In this case, no sharp MMT is observed. Each MR isotherm is dominated by a broad maximum that moves towards stronger fields with rising temperature, and it is present also in the paramagnetic state. Furthermore, it is worth mentioning that in this field geometry the absolute value of MR is significantly reduced as compared to the other field orientation. This observation also highlights strong anisotropic nature of the MR data. In addition, solely in the data collected at $T$ = 0.7~K, a fairly smeared peak is seen near $\mu_0H$ = 8~T. This feature may signal a kind of spin reorientation in strong magnetic fields, actual nature of which remains unsolved. Nevertheless, comparison of the MR results obtained for different field orientations confirms that the magnetic easy direction in Ce$_3$PdIn$_{11}$ is most likely parallel to the crystallographic $c$ axis.

\subsection{Magnetic phase diagram}

Summarizing the results obtained from the heat capacity and magnetotransport measurements performed with external magnetic field applied along the crystallographic $c$ axis, we constructed a magnetic phase diagram presented in  Fig. 6a which is in concert with that reported previously \cite{Kratochv} (in order to highlight the field evolution of the heat capacity singularities, we presented this phase diagram in a form of color contour plot). It is quite evident from Fig. 6a that the phase diagram can be divided into three separate regions. Initially, both $T_{\rm {N1}}$ and  $T_{\rm {N2}}$ decrease with increasing the magnetic field strength. Then, both critical temperatures merge into a single transition at $T_{\rm {M}}$. Finally, in the third segment of the phase diagram, the ordered state again comprises two separate features at $T_{\rm {M1}}$ and $T_{\rm {M2}}$. Interestingly, the heat capacity singularity associated with $T_{\rm {M2}}$ is very sharp and pronounced as can be clearly inferred from the contour map. This behavior signals possible unconventional nature of the transition which may imply a significant reconstruction of the Fermi surface, thus provoking the idea of field induced Lifshitz transition in Ce$_3$PdIn$_{11}$. This peerless feature demands further detailed experimental studies exploring the Fermi surface geometry. In addition to these phase boundaries, there exists another one (see Fig. 6a), which can be attributed to MMT, clearly observed in the MR data. The latter feature is hardly dependent on the magnetic field strength, and seems to have a first-order character (see above). The magnetic phase diagram of Ce$_3$PdIn$_{11}$ featuring both first and second order type transitions, is fairly similar to that derived before for the Pt-bearing counterpart Ce$_3$PtIn$_{11}$ \cite{Das}. 

\begin{figure}[htb!]
	\includegraphics[height=11cm, width=9cm, keepaspectratio]{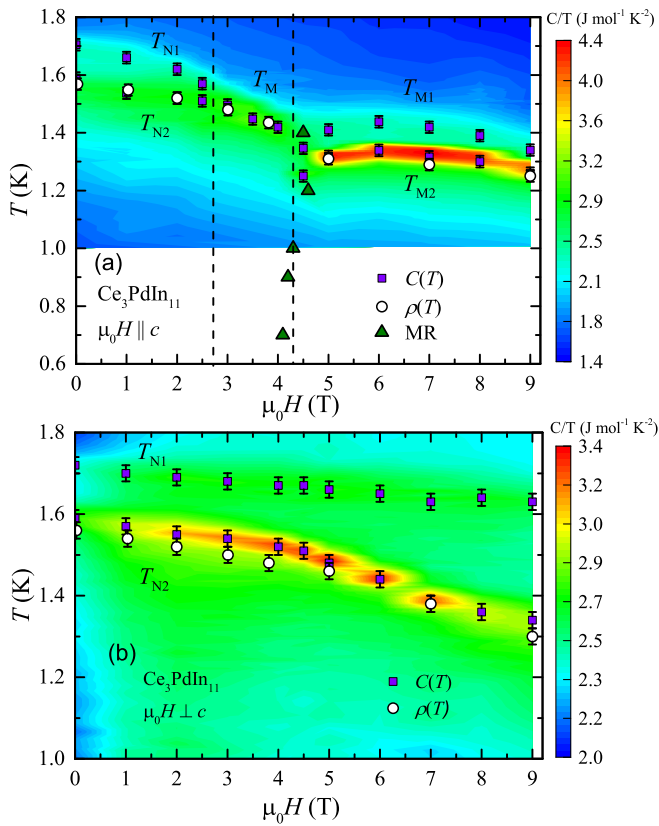}
	\caption{\label{fig:Phase}(Color online)  Magnetic phase diagram (presented as color contour) of single-crystalline Ce$_3$PdIn$_{11}$ derived from the heat capacity and electrical transport data taken with magnetic field directed (a) along the crystallographic $c$ axis and (b) perpendicular to the $c$ axis. The vertical dotted lines in panel (a) serve as a guide to the eye sectioning the phase diagram into three different regions, as discussed in the text.}
\end{figure}

Fig. 6b, corresponds to the $\mu_0H$-$T$ phase diagram constructed for Ce$_3$PdIn$_{11}$ based on the heat capacity and magnetotransport data collected for $\mu_0H \perp c$-axis. In contrast to the afore-discussed field orientation, here $T_{\rm {N1}}$ and $T_{\rm {N2}}$ do not merge, and their separation systematically increases with increasing field up to 9~T. Thus, it is essential to perform neutron diffraction measurements to explore the magnetic structure of Ce$_3$PdIn$_{11}$.

\section{Conclusions}

In conclusion, the results of our detailed investigation of the bulk physical properties on single-crystalline Ce$_3$PdIn$_{11}$ elucidated two AFM phase transitions at $T_{\rm {N1}}$ = 1.68~K and $T_{\rm {N2}}$ = 1.56~K, and superconductivity below $T_{\rm c}$ = 0.58~K. Observation of such coexistence of AFM and superconductivity in  Ce$_3$PdIn$_{11}$  is in line with the previous work by M. Kratochv\'{i}lov$\acute{\textup{a}}$ $et~ al.$ \cite{Kratochv}. The absence of any hysteresis in $C(T)/T$ for magnetic fields $\ge$ 3.5~T applied parallel to the crystallographic $c$ axis does not support the scenario of first--order character of the AFM transitions in strong magnetic fields, suggested in Ref.\cite{Kratochv}. However, the overall shapes of the associated singularities in the heat capacity hint at some unconventional second--order type transitions, which demand further detailed investigations. The MR data collected with external magnetic field applied parallel to the crystallographic $c$ axis clearly revealed the metamagnetic behavior. In contrast, no similar features were observed when the applied field was perpendicular to $c$ axis. This observation elucidates that the magnetic easy direction in Ce$_3$PdIn$_{11}$ is parallel to the crystallographic $c$ axis. Our heat capacity and electrical transport measurements carried out in different magnetic fields applied along and perpendicular to the $c$-axis conjointly manifested strongly anisotropic influence of magnetic field on the AFM state. In a manner similar to that reported before for Ce$_3$PdIn$_{11}$ \cite{Kratochv} and Ce$_3$PtIn$_{11}$ \cite{Das}, for $\mu_0H\parallel c$-axis, the two AFM phase boundaries first merge and then split again on increasing the magnetic field strength. In turn, for $\mu_0H\perp c$-axis, $T_{\rm {N1}}$ and $T_{\rm {N2}}$ remain separated up to the strongest field studied. Further comprehensive investigations involving neutron diffraction and muons spin rotation spectroscopy have been envisaged to address in more details the actual nature of the different magnetically ordered states in Ce$_3$PdIn$_{11}$. In addition, the most intriguing issue of the coexistence in this material of the long-range magnetic ordering and the superconductivity will be tackled in our forthcoming studies.

\section{Acknowledgment}
The authors are indebted to Dr. Marek Daszkiewicz for the crystal structure characterization. This work was supported by the National Science Centre (Poland) under research grant No. 2015/19/B/ST3/03158.

\end{document}